\documentclass[fleqn,10pt]{wlscirep}
\usepackage{amssymb}
\usepackage[utf8]{inputenc}

\title{Spin-Wave Optics in YIG by Ion-Beam Irradiation}
\author[1,*]{Martina Kiechle}
\author[2]{Adam Papp}
\author[1]{Simon Mendisch}
\author[1]{Valentin Ahrens}
\author[1]{Matthias Golibrzuch}
\author[3]{Gary H. Bernstein}
\author[3]{Wolfgang Porod}
\author[2]{Gyorgy Csaba}
\author[1,+]{Markus Becherer}
\affil[1]{Department of Electrical and Computer Engineering, Technical University of Munich, Germany}
\affil[2]{Faculty of Information Technology and Bionics, P\'{a}zm\'{a}ny P\'{e}ter Catholic University, Budapest, Hungary}
\affil[3]{Department of Electrical Engineering, University of Notre Dame, Notre Dame, IN, 46556}
\affil[*]{martina.kiechle@tum.de}
\begin{abstract}
We demonstrate direct focused ion beam (FIB) writing as an enabling technology for realizing spin-wave-optics devices. It is shown that ion-beam irradiation changes the characteristics of YIG films on a submicron scale in a highly controlled way, allowing to engineer the magnonic index of refraction adapted to desired applications. This technique does not physically remove material, and allows rapid fabrication of high-quality architectures of modified magnetization in magnonic media with minimal edge damage (compared to more common techniques such as etching or milling). 
By experimentally showing magnonic versions of a number of optical devices (lenses, gratings, Fourier-domain processors) we envision this technology as the gateway to building magnonic computing devices that rival their optical counterparts in their complexity and computational power. 

\end{abstract}
\begin{document}
\flushbottom
\maketitle
\thispagestyle{empty}
\section{Introduction}
\label{Introduction}
A major motivation behind magnonics research is to replicate the functionality of optical devices in chip-scale devices that are amenable to integration with microelectronic circuitry \cite{csaba2014spin}$^,$\cite{csaba2017perspectives}. This way, the functionality of coherent optical computers could also be cloned in the magnonic domain. Spin waves (magnons) display interference phenomena that resemble the ones shown by optical waves, but they offer submicron (possibly sub-100\,nm) wavelength and can be launched and detected by electrical means. Chip-scale optically-inspired devices also provide a pathway to much needed energy-efficient neuromorphic and edge-AI computing components \cite{wang2021inverse}$^,$\cite{papp2021nanoscale}.
\\
It has, however, remained elusive to produce spin-wave optics that approach the 'ideal' behavior of optical components. This is largely due to the fact that no working technology is available to control the propagation characteristics of spin waves to the extent that is possible in optics. Ideally, one would want to realize any fine-grained spatial distribution of the index of refraction, as this provides a high degree of freedom in device design.
Magnonic devices are almost exclusively made from Yttrium Iron Garnet (YIG) substrates, having low attenuation that enables propagation over long distances. Lithographic patterning of YIG allows defining some spin-wave optic functions\cite{papp2017nanoscale} but has technological challenges \cite{zhu2017patterned}$^,$\cite{trempler2020integration}, as etched or milled film edges introduce undesired behaviors\cite{schlitz2019focused}. Even a well-controlled YIG patterning technology would be insufficient to replicate the propagation of electromagnetic waves, which propagate in vacuum, and optical devices can be made by patterning a transparent material to an appropriate shape. Essentially, refractive spin-wave optics need additional materials or intrinsic YIG film modifications, as magnons do not propagate in air. 
\\
A few pathways have been proposed to realize engineered YIG substrates. It is possible to use localized magnetic fields to steer spin waves \cite{chumak2015magnon}, but this requires a second magnetic layer to generate the fields and arbitrarily-shaped field profiles cannot be realized. Previous work has shown the use of heat distributions \cite{vogel2020optical} to generate refraction index profiles, a solution likely impractical in chip-scale devices. Local exchange bias may also be used for magnonic optics\cite{albisetti2017nanopatterning}, but so far this works only on metallic systems, which are high-damping in nature.
\\
In multilayered magnetic systems, it is well established that focused ion beam (FIB) irradiation modifies the magnetic properties \cite{chappert1998planar}$^,$\cite{mendsich2020perpendicular} without actually removing material or creating edges. The FIB irradiation affects magnetic properties on a sub-50 nm size scale, which is a resolution that is hardly achievable by the combination of lithography and etching.
The effect of FIB on magnetic multilayers motivated our work to study the effects of direct FIB-ing on the magnetic properties of YIG. Using 50 keV Ga$^+$ ions, the applied ion doses are chosen to not physically remove material, but to implant Ga$^+$ into the YIG thin film and locally alter the crystalline structure.
\\
In our work, we first demonstrate the effect of FIB irradiation on spin waves in YIG for plane-wave propagation and characterize the dependence of magnonic wavelength on FIB dose,  obtaining $n$. Lenses and diffraction gratings are designed with a binary irradiation pattern -- in a similar fashion to elementary optical devices, where light propagates either through glasses or vacuum.
We find that the effect of FIB irradiation on YIG films can be modeled precisely enough by assuming an effective magnetization ($M_\mathrm{eff}$) value that varies with the FIB dose, which is in agreement with the findings in other work \cite{ruane2018controlling}. The spatially varying $M_\mathrm{eff}$ results in a spatially varying dispersion relation for the film, which, in turn, may be understood as a spatially varying index of refraction $n$. As a consequence, FIB irradiation allows the realization of (almost) arbitrary two-dimensional $n$ profiles.
Going beyond binary patterns, we demonstrate that FIB irradiation is especially useful for graded-index magnonics\cite{papp2016optically}$^,$\cite{davies2015towards} as it allows a continuous and high-resolution variation of $n$ across the film surface. As a highly meaningful example of such systems, a $\mathrm{4f}$ Fourier-domain signal processor will be shown in Section \ref{sec:results}. The demonstration of a $\mathrm{4f}$ system opens the door to the realization of a variety of optically-inspired computing systems \cite{chang2018hybrid} using magnons.
We envision that the FIB technology shown here will readily provide access to magnonic devices that may rival on-chip optics in their functionality, and consequently act as the gateway to magnonic integrated circuits (in analogy to photonic ICs). While spin waves have limitations (such as damping that should be compensated by some amplification mechanism for large-scale devices), they have benefits (such as nonlinearities \cite{bauer2015nonlinear}$^,$\cite{papp2021nanoscale}) that open up applications unreachable for photonic ICs. 
\section{Methods}
\label{sec:methods}
\begin{figure}[ht]
    \centering
    \includegraphics[width=\textwidth]{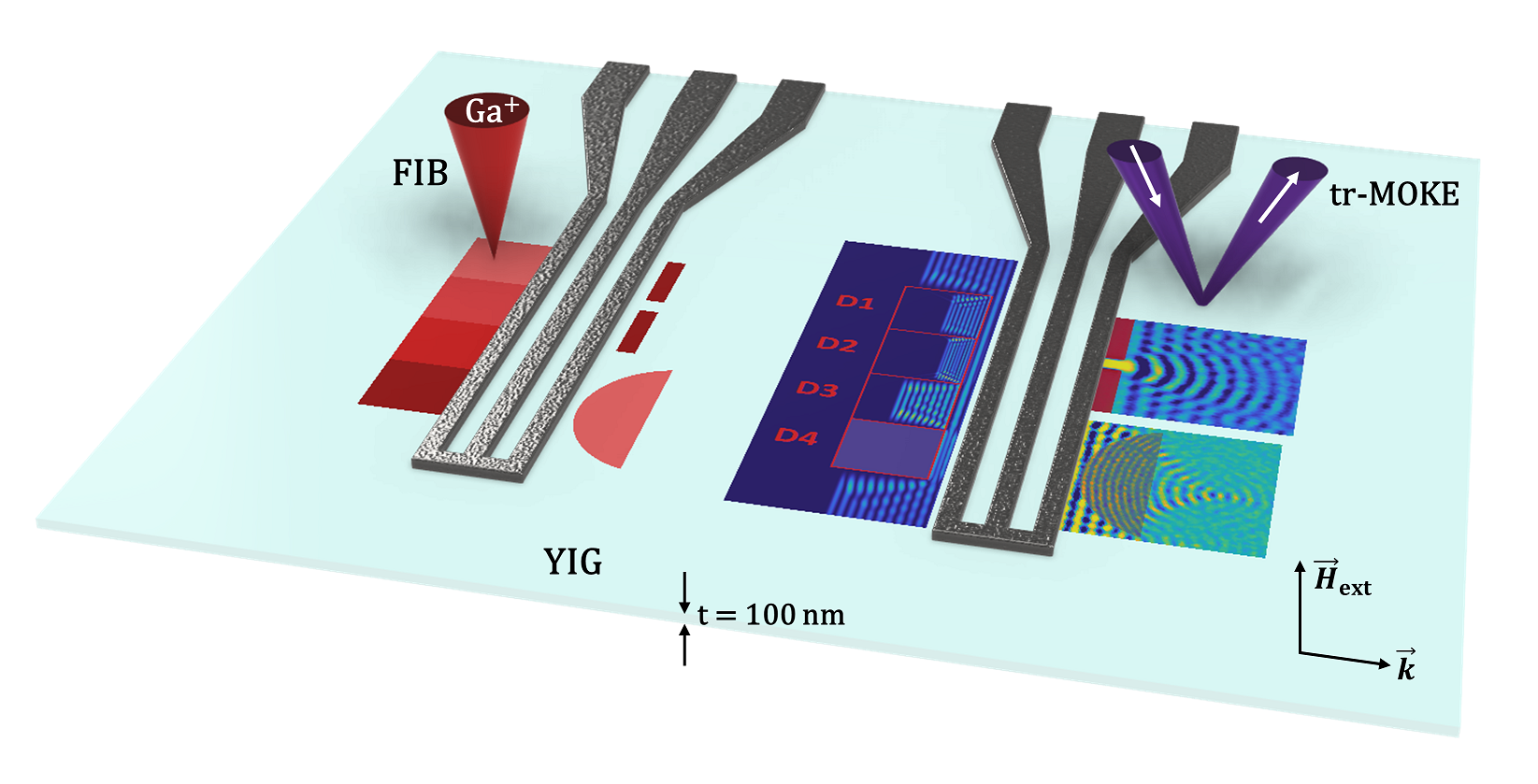}
    \caption{Representation of the experimental arrangement. Left part: Areas in YIG next to the excitation antenna are directly irradiated with FIB at different ion dose levels (indicated by the intensity of the red color), and an ion dose dependent change of $M_\mathrm{eff}$ is found. This is used for the application examples: a lens at a low dose with a modified $n$ and a single slit at a high dose. Right part: Spin wave propagation patterns in the FIB-irradiated regions are imaged with trMOKE.}
    \label{fig:3dsketch}
\end{figure}

The effect of FIB irradiation is characterized by recording the spin-wave (SW) waveform using longitudinal time-resolved magneto-optical Kerr effect microscopy (trMOKE) and determining the SW wavelength change at various ion dose levels. Fig.~\ref{fig:3dsketch} gives an overview of the experimental techniques used for fabrication and metrology. An in-house sputter-deposited YIG thin film ($t_\mathrm{YIG}$ = 100 nm) with co-planar microwave antennas ($\mathrm{s_{CPW}}$, $\mathrm{g_{CPW}}$\,=\,2-5 $\mu$m) is bonded to a PCB board from where it is fed with a microwave signal. Areas next to the excitation antennas are FIB-irradiated at different ion doses and shapes. Fabrication details can be found in Sec. \ref{supplementary}. Subsequently, 2D spin-wave patterns are imaged with a longitudinal, time-resolved Kerr microscope in forward volume configuration. 
\\
In the home-built trMOKE apparatus (for details, see Supplementary Sec. \ref{sec:trMOKE}) the film is magnetized out-of-plane (along $M_z$) and the dynamic $m_x$ and $m_y$ components display wave propagation. The FVSW mode is isotropic and therefore shows the closest analogy to optical wave propagation. The spatial resolution is diffraction-limited at about $d=$~0.4~$\mu$m, limiting the shortest detectable spin-wave wavelength to about $\lambda=$~1~$\mu$m.
\subsection*{Characterizing the effect of FIB irradiation in YIG}
\label{subsec:FIBeffect}
\begin{figure}[ht]
    \centering
    \includegraphics[width=\textwidth]{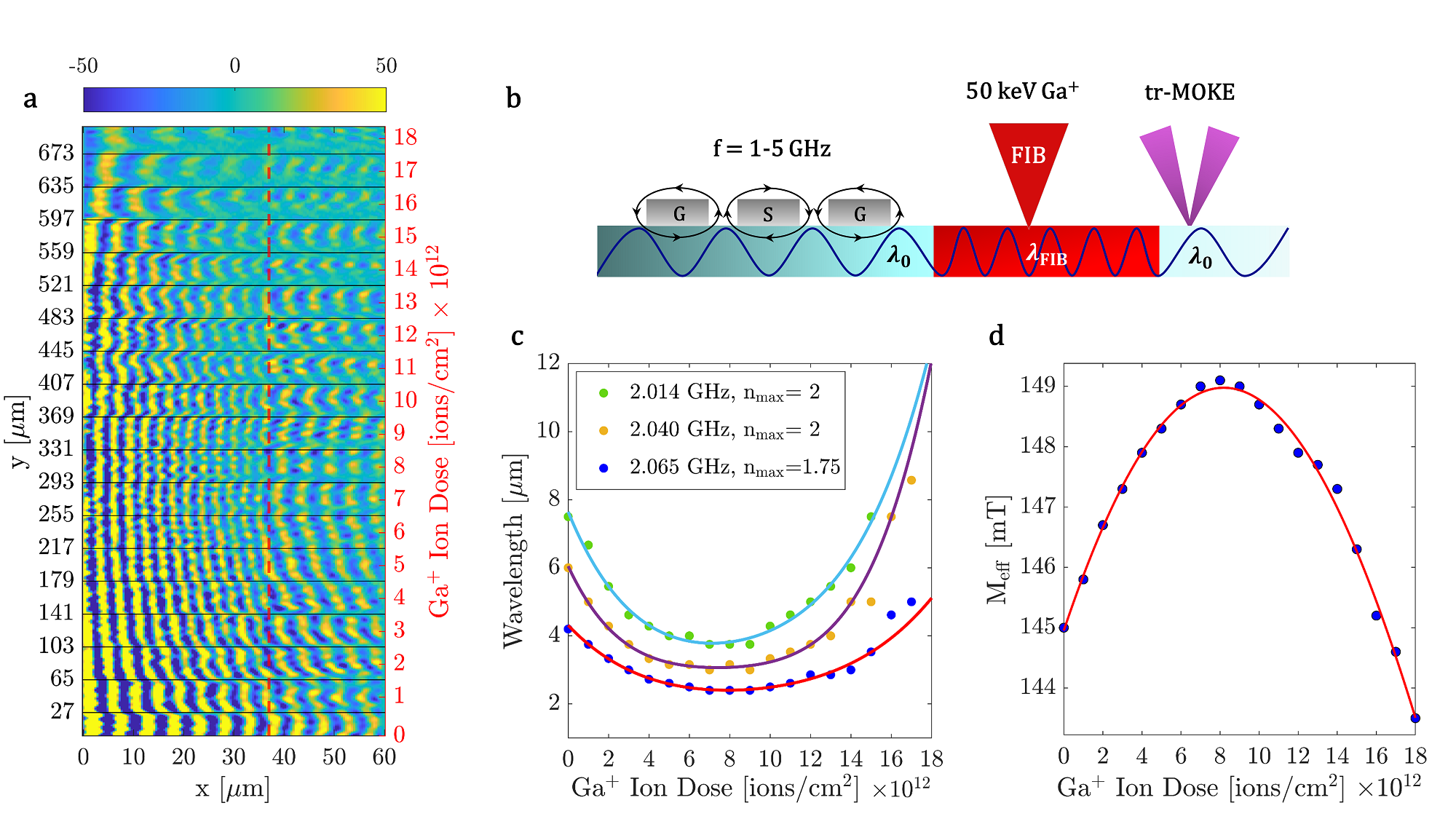}
    \caption{Illustration of the ion dose dependent change of $M_\mathrm{eff}$ in YIG. (a) trMOKE image of coherently excited spin waves in regions of linearly increasing ion doses. The bottom line shows the actually excited wavelength in unirradiated YIG. (b) 1D sketch of the experiment: The area next to the excitation antenna is FIB irradiated, whereby the magnetization and hence the wavelength is locally changed. (c) Wavelength change vs. ion dose profile at different frequency settings with an external field of 214 mT. (d) Ion dose dependent change of $M_\mathrm{eff}$, numerically calculated from dispersion relation.}
    \label{fig:dosemap}
\end{figure}

50 keV accelerated Ga$^+$ ions are used to irradiate YIG thin films at ion doses ranging from $1\cdot10^{12}$ to $1\cdot10^{15}$ ions/cm$^2$, with the purpose of manipulating the magnetic properties locally. As a dose calibration method, regions with linearly increasing ion doses are irradiated next to the excitation antenna and the SW wavelength change due to a $M_\mathrm{eff}$ modification is measured. Fig.~\ref{fig:dosemap} shows the resulting wavelength profiles vs. the applied ion dose. 
The degree of change in the magnetic properties is dependent on the applied Ga$^+$ ion dose and also on the acceleration voltage if the film thickness is larger than the ion penetration depth. The SRIM\cite{ziegler2013srim} simulated mean ion implantation depth of Ga$^+$ in a 100\,nm thick YIG film is 24\,nm and makes about a third of the total thickness (considering the effective thickness due to e.g. Ga diffusion into the first few layers, sputter process imposed inhomogenities, etc.). Nonetheless, we characterize how spin-wave propagation is affected in the effective layer and therefore across the entire film thickness. Insights into FIB-irradiation-induced changes of YIG on the crystal level can be found in Sec.~\ref{supplementary}. 
Interpreting the results in terms of micromagnetic parameters, we find that FIB-ing can be accurately modeled in terms of modifying the effective magnetization $M_\mathrm{eff}$ and the magnetic damping $\alpha$. A low ion dose (up to $6\cdot10^{12}$ ions/cm$^2$) increases $M_\mathrm{eff}$, and hence decreases the wavelength, while $\alpha$ increases only moderately. For higher ion doses, a turning point is reached and $M_\mathrm{eff}$ decreases again, while magnetic damping increases until spin-wave propagation is inhibited (doses larger than $1\cdot10^{14}$ ions/cm$^3$). The wavelength vs. ion dose profile is non-linear and wavelength-dependent due to the inherent non-linear dispersion relation of spin waves. In the low-dose regime, $M_\mathrm{eff}$ is slightly increased, which is used for the demonstration of optical elements with a singular refractive index in analogy to glasses.\\
Applications of higher complexity, such as continuous refractive media, can be modeled by magnetization landscapes. Using FIB, this means changing the ion dose point by point. Alternatively, it is also possible to change the filling factor in pixel space and set the value of $M_\mathrm{eff}$ this way. This way, only a single ion dose has to be applied, and the average magnetization is changed due to a density gradient. We use a pixel size of 40 nm in the experiments, which is about two orders of magnitude smaller than the applied spin-wave wavelength. This technique (demonstrated in Sec.~\ref{sec:GRIN}) is well suited for continuously changing wave propagation properties, as is done with graded-index (GRIN) optics \cite{davies2015towards}. 
A more trivial (but often-needed) use of FIB irradiation is to apply a high dose that entirely blocks propagation, and hence reflects magnons. The high dose effectively destroys the magnetic properties ($M_\mathrm{eff}$=0) such as shown in \cite{papp2021experimental}. 
\subsection*{Tuning the magnonic index of refraction and saturation magnetization by FIB}
\label{sec:refractive_index}
In order to design spin-wave 'replicas' of optical devices, it is instructive to define the magnonic index of refraction, $n$. The relative change of the magnonic refractive index is extracted from the wavelength change in the untreated vs. the irradiated YIG film part ($\lambda_0$ vs. $\lambda_{FIB}$):
\begin{equation} 
n_\mathrm{rel} = \lambda_0 /\lambda_{FIB}
\label{equ:neff}
\end{equation}
The highest achievable refractive index  $n_\mathrm{max} = \lambda_0 /\lambda_\mathrm{min}$ corresponds to the ion dose that generates the highest $M_\mathrm{eff}$, and hence the smallest wavelength $\lambda_\mathrm{min}$ with respect to the initial parameters. In order to model the correlation of FIB-ing and the magnetic properties, the effect of FIB is best understood as an ion dose dependent change of the effective magnetization $\Delta M_\mathrm{eff}=M_\mathrm{eff,FIB}-M_\mathrm{eff,0}$. The results are shown in Fig.~\ref{fig:dosemap}d (see previous section for details).
Due to the highly nonlinear nature of the magnonic dispersion relation, $n_\mathrm{eff}$ is only valid for a certain spin-wave frequency $f$ and the corresponding $\lambda_0$ wavelength. We target wavelengths $\lambda_0$ that can be efficiently excited and detected by coplanar waveguide antennas in use. Generally, the refractive index for a specific wavelength can be calculated by numerically solving the dispersion relation for $k$-vectors at the initial $M_\mathrm{eff,0}$ and for $M_\mathrm{eff,FIB}$ at the respectively chosen ion dose:
\begin{equation} 
n_\mathrm{rel} =  k(M_\mathrm{eff,0},f,H_\mathrm{ext},t,A_\mathrm{exch})/k(M_\mathrm{eff,FIB},f,H_\mathrm{ext},t,A_\mathrm{exch})
\label{equ:neff_disp}
\end{equation}
In Eq.~\ref{equ:neff_disp}, $f$ is the microwave frequency used for excitation, $H_\mathrm{ext}$ the applied bias field normal to the film plane (forward volume), $t$ is the film thickness, and $A_\mathrm{exch}$ is the exchange stiffness. 
\section{Results}
\label{sec:results}
\subsection*{Design and fabrication of optically inspired magnonic elements}
\begin{figure}[ht]
    \centering
    \includegraphics[width=\textwidth]{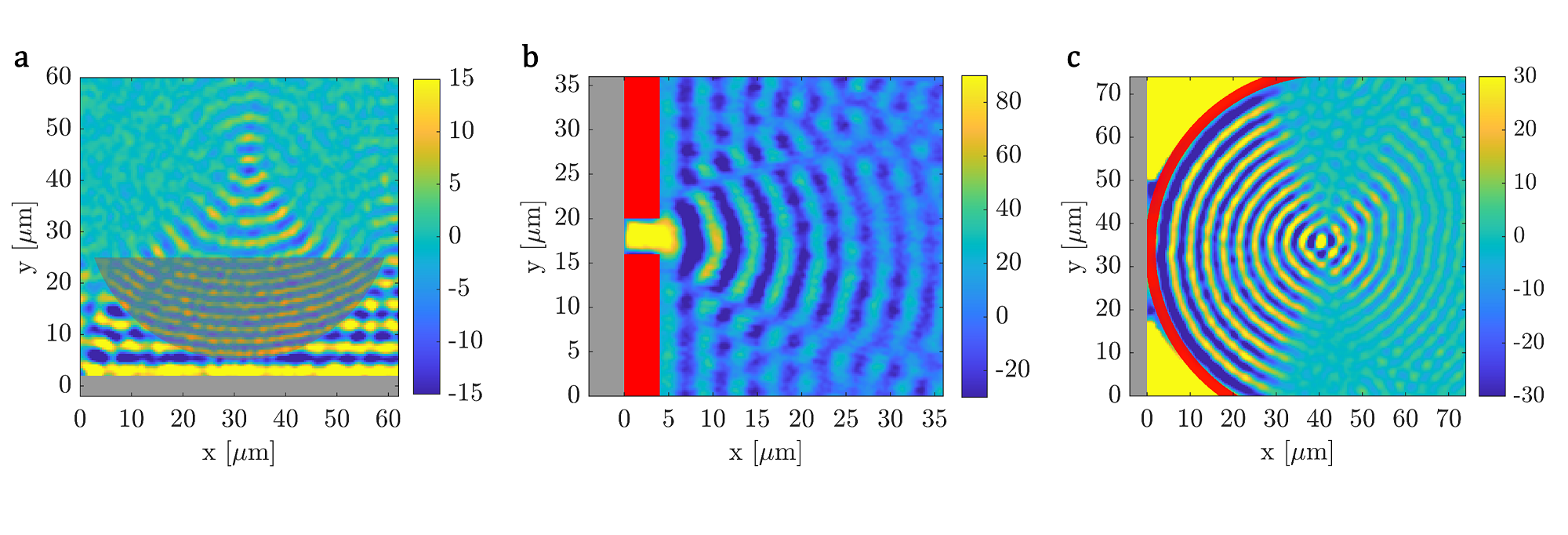}
    \vspace*{-10mm}\caption{Demonstration of refractive, diffractive and reflective optical components fabricated with FIB irradiation. (a) D-shaped spherical lens with a curvature radius of 30 $\mu$m and a clearly shortened wavelength in the inside (semi-transparent red, n about 1.8). (b) An optical slit (4 $\mu$m width). (c) A semicircular source (R=40 $\mu$m) with spin waves excited through the high-amplitude resonance behind the circle. The high-amplitude region appears as a saturated yellow region. Red lines indicate a high FIB dose that blocks spin wave propagation.}
    \label{fig:lens_slit_circle}
\end{figure}

\begin{figure}[ht]
    \centering
    \includegraphics[width=\textwidth]{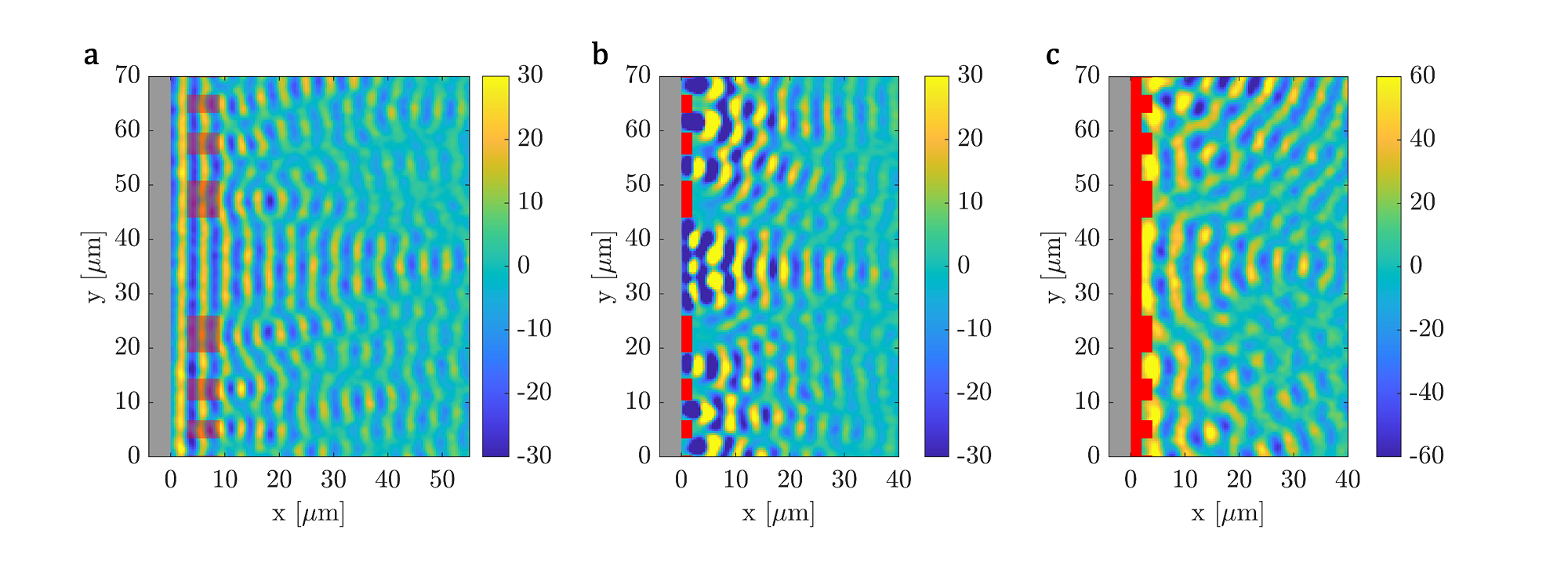}
    \vspace*{-5mm}\caption{Spin-wave diffraction and focusing in FIB-treated regions (red areas) by reference to an optical (Fresnel) zone plate. Different ion doses trigger distinct operating mechanisms. (a) Phase-shift-induced focusing achieved with a low ion dose. The shortened wavelength in the FIB areas exits the zones with a phase shift close to 180$^{\circ}$. (b) A high ion dose that causes a local barrier for spin waves, analogously to an optically opaque region. (c) A blocking wall before the actual zone plate in combination with a high excitation amplitude excites spin waves via dipole field coupling that accumulate on the back side.}
    \label{fig:zoneplates}
\end{figure}

To replicate the behavior of conventional optical elements (i.e. glasses), one may utilize only one other $n_\mathrm{eff}$ value in addition to that of intrinsic YIG (where $n_\mathrm{rel}$-1). Fig.~\ref{fig:lens_slit_circle}a shows a trMOKE image of a plano-convex lens realized with this binary technique.
The Lensmaker's equation\cite{lenses2008} is used for the given refractive index of 1.8 at a dose of 7$\cdot$10$^{12}$ ions/cm$^2$. This is the maximum $n_\mathrm{eff}$ change achievable at this wavelength. The lens has a curvature radius of 30 $\mu$m and thickness of 20 $\mu$m, resulting in a focal distance of about 37 $\mu$m. The trMOKE measured image of the lens reveals parameters closely matching the design target calculated from the optical formulas. 
A different way of using the FIB irradiation is shown in Fig.~\ref{fig:lens_slit_circle}b, where we show single-slit spin-wave diffraction achieved by locally eliminating the magnetization in the red areas through irradiation at a high ion dose of 1$\cdot$10$^{15}$  ions/cm$^2$. The diffraction pattern closely matches textbook images of optical diffraction for a single slit, and a plane wave couples through from behind the FIB irradiated part. To complete the parallels with elementary optical components, Fig.~\ref{fig:lens_slit_circle}c demonstrates a circular-shaped source focusing spin waves at a distance that equals the radius.
The curved surface acts as a secondary spin wave source, where spin precession is driven by the high-amplitude, spatially uniform oscillations of the area between its spherical backside and the excitation antenna. 
The primary non-linear precession excites linear spin waves on the inside of the sphere via dipole coupling, a mechanism explored in \cite{papp2021experimental}.
\\ Another key function in optics is the ability to phase shift, and we show this feature by the example of a Fresnel phase plate in Fig.~\ref{fig:zoneplates}a. 
The focusing effect occurs via constructive interference of beams propagating with respect to a 180$^{\circ}$ phase difference to each other. This phase shift is achieved by the change of $n$ over the 6 $\mu$m length of the zones (semi-transparent red overlays) so that the 180$^{\circ}$ difference occurs at the zone plate's exit plane. Alternatively to this phase-shift-based zone plate, a Fresnel zone plate is realized by simply blocking waves in the regions where they would destructively interfere at the desired focal point. In this device (Fig.~\ref{fig:zoneplates}b) the phase shifters are replaced by highly-irradiated regions (red lines). A third zone-plate demonstration uses the effect we also exploited in the circularly radiused source in Fig.~\ref{fig:lens_slit_circle}c, i.e. that a coherent wavefront is created at the boundary of a high-dose region. By shifting the wavefront about $\lambda/2$ between the zones, the desired focusing effect is achieved, as shown in Fig.~\ref{fig:zoneplates}c). The geometrical arrangement of the zones is chosen to focus a 4 $\mu$m wavelength at a distance of 40 $\mu$m from the exit plane. Since the excited wavelength is a little different in a), b) and c), the observed focal distance varies accordingly. Diffractive devices demonstrate the high resolution of FIB patterning and serve as proof that little damage is done outside the irradiated area in the YIG films.
\subsection{Gradient index and Fourier optics for spin waves}
\label{sec:GRIN}
\begin{figure}[ht]
    \centering
    \vspace*{-0.2cm}
    \includegraphics[width=\textwidth]{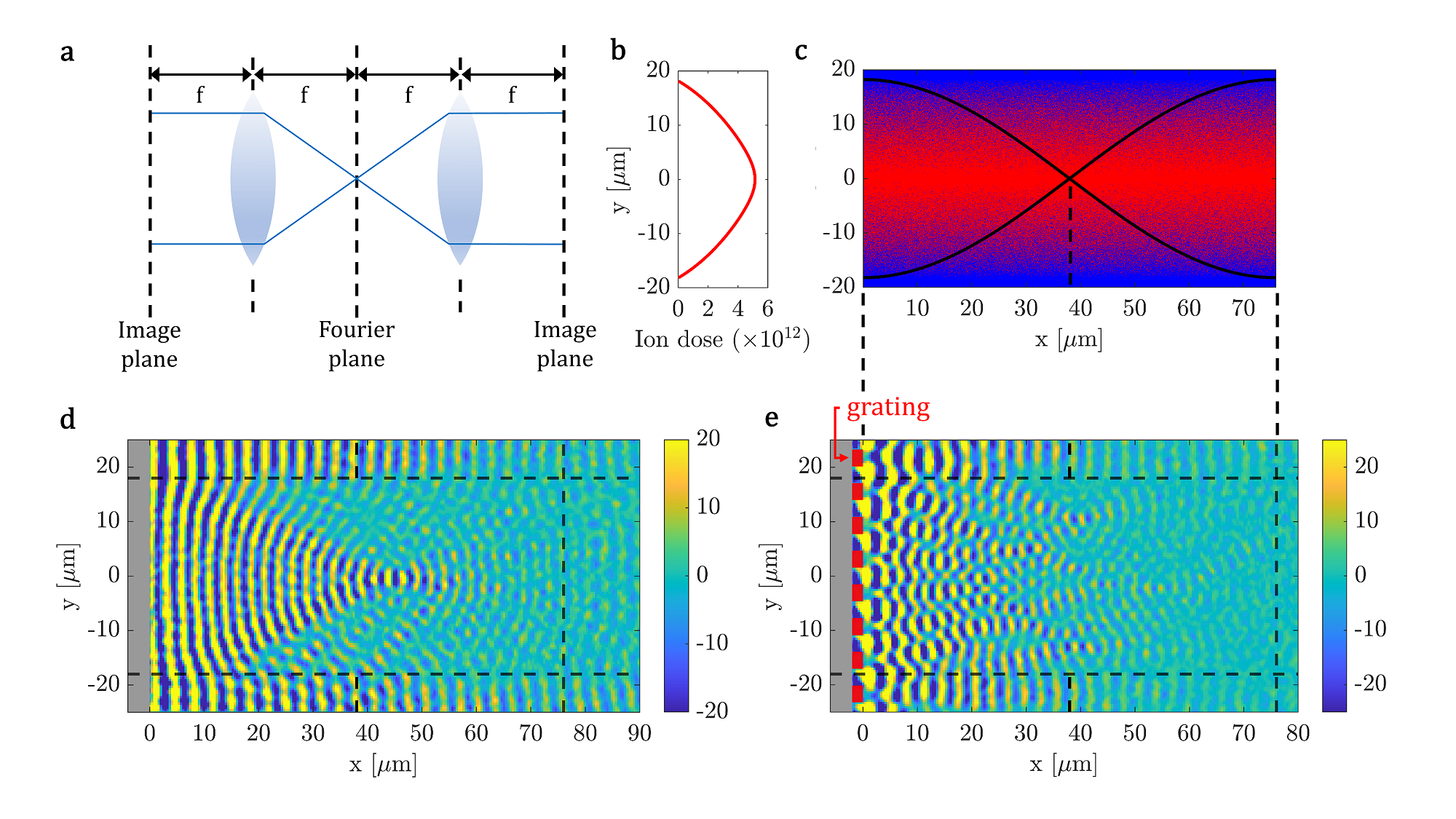}
    \vspace*{-0.7cm}\caption{Demonstration of a 4$\mathrm{f}$ system for spin waves realized with a FIB irradiated graded index magnetization. (a) Working principle of a conventional $\mathrm{4f}$ system based on two consecutive lenses. (b) 2D irradiation pattern of a full pitch GRIN lens realized by a filling factor variation in the FIB image. The 1D ion dose profile right beside is calculated from the measured wavelength change vs. ion dose profile. (c) trMOKE image of the GRIN lens irradiated in YIG. (d) A diffraction grating is irradiated at the anterior plane. The system images the first order diffraction at 30$^\circ$ on the Fourier plane.}
    \label{fig:GRIN}
\end{figure}

In Fourier optics \cite{ambs2010optical}$^,$\cite{goodman2005introduction}, linear processing functions mostly rely on the Fourier transform property of lenses - easily moving between the real and the Fourier domains enables a number of signal processing primitives. Similarly, Fourier-optics devices for spin waves could serve as building blocks for useful computing functions. 
\\
Perhaps the most illustrative of Fourier optics devices is the 4f system illustrated in Fig. \ref{fig:GRIN}a. The image (wave amplitude and phase on the image plane) is Fourier transformed by the first lens and this Fourier transform appears in the Fourier plane, which is inverse-Fourier transformed by the second lens. Any Fourier-domain manipulation of the image (such as filtering, convolution, matched filtering) can be accomplished by a filter placed in the Fourier plane that alters the magnitude and/or phase of the spectral components of the image. 
\\
While Fourier optics components could be put together from concave lenses such as the one shown in Fig. \ref{fig:lens_slit_circle}a), the lens boundaries introduce undesired reflections and diffraction effects, which can largely be avoided in graded-index (GRIN) optics \cite{davies2015graded}. Since FIB irradiation can continuously tune $n_\mathrm{eff}$ of a magnetic film by changing the filling factor of an image in pixel space, we can create a magnetization landscape of arbitrary shape, including a GRIN lens profile.  
\\
To produce a refractive index gradient of a certain shape, the ion-dose profile needs to be determined for the desired wavelength (or index of refraction) profile. Here we used the measurements from Fig. \ref{fig:dosemap}c. In case of a GRIN lens, the wavelength profile for a parabolic refractive index change can be written as\cite{Smith2007modern}
\begin{equation} 
\lambda =  \frac{\lambda_{FIB}}{1-0.5(2\pi y/4f)^2}.
\label{equ:lambda_GRIN}
\end{equation}
For the calculation of the required ion dose profile that results in the desired magnetization gradient, we use the ion dose vs. wavelength profile from Fig. \ref{fig:dosemap}c and insert the inverted version into the GRIN lens wavelength profile (Equ.~\ref{equ:lambda_GRIN}), resulting in the ion dose profile in Fig.~\ref{fig:GRIN}b. This profile is used for the density distribution of the FIB image in pixel space, and the 2D irradiation pattern of a 4f GRIN lens with a diameter of 18.48 $\mu$m and a length of 76 $\mu$m is shown in Fig. \ref{fig:GRIN}c. Experimentally, this image is irradiated at a peak ion dose of 5.2$\cdot$10$^{12}$, resulting in respectively lower doses across the diameter due to the density variation. 
The measured spin-wave profile of the fabricated GRIN lens performing a Fourier transform of the plane wave excited at the microwave antenna right behind the lens is shown in Fig. \ref{fig:GRIN}d. The focal distance is slightly longer than calculated (45 $\mu$m instead of 38 $\mu$m), that is most likely a result of a bias field inaccuracy or a deviation from the desired ion dose. 
\\
To demonstrate FT properties of this GRIN lens, we irradiated a diffraction grating in front of the GRIN lens entrance plane by using a high dose, as shown in the slit experiment in Fig.~\ref{fig:lens_slit_circle}b. The grating is designed to have a diffraction angle of 30$^\circ$ for $\lambda_0$=3.1 $\mu$m wavelength (grating constant = $\lambda_0$=6.2 $\mu$m, thickness = 2 $\mu$m). The expected Fourier plane (FP) should occur at 38 $\mu$m, and we observe a slight deviation that has also occurred in the focal distance when we tested the GRIN lens with a plane wave. The occurring focal points left and right of the center peak in the FP represent the first order diffraction of the grating, and correspond to the expected diffraction angle. 
\section{Discussion}
\label{discussion}
We can group the presented devices based on multiple properties. First of all, a part of the devices in this paper are refractive, i.e. ion irradiation is used for changing the effective refractive index in the specified region, while others are reflective or diffractive, i.e. relatively high doses are used to produce a perimeter around the device geometry. Here, the saturation magnetization in the YIG film changes abruptly -- practically plunging to zero. In the former case, spin waves propagate through the irradiated region (although with modified dispersion), while in the latter case spin waves are blocked, reflected, or generated on the boundaries between the intrinsic and irradiated YIG regions. This brings us to the next distinction based on whether the device is used for manipulating an incident wavefront or if it's used to indirectly excite a wavefront with a sophisticated shape. From the perspective of functionality, we have demonstrated elements for focusing waves and also diffraction gratings (the zone plates being a combination of the two). Inarguably, the most sophisticated demonstration we present here is the GRIN lens and the 4f system based on it, which contains a pointwise varying refractive irradiation that is used for focusing waves, combined with a diffraction grating that block parts of the incident wavefront in a periodic manner, transmitting only a single spatial-frequency component. The above distinctions provide justification of the rich selection of demonstrations in this paper, representing the flexibility and wide applicability of a single fabrication technique to implement a wide range of optical elements. This set of devices can be considered complete in the sense that a full linear signal processor might be realized based on them, as demonstrated through the 4f system.
\\
The refractive index modification for spin waves in YIG through a FIB induced magnetization change can be accurately calculated from the ion-dose-dependent wavelength change. With this, we are able to design devices not only where binary refractive arrangements are needed, but also create smooth magnetization transitions (gradients) that are essential in GRIN optics. With the strategy to use a halftoning technique (stochastical filling factor in pixel space) on the FIB image to produce an ion-dose gradient, one global ion dose can be used for the entire image, eliminating the need for multiple irradiation steps. Since the spotsize of the FIB is two orders of magnitude smaller than the spin-wave wavelength, this simplification is not expected to degrade the device performance.
\\
As the complexity of the 4f system is the highest among the presented devices, the limitations of the technology (spin waves and also our facilities) is the most evident here. As it is visible in Fig. \ref{fig:GRIN}\textbf{d} and \textbf{e}, damping limits the useful length of the GRIN lens to tens of wavelengths. We presume that our demonstration is not optimal from this perspective in multiple ways: with better quality (homogeneity and damping) YIG films, and optimization of parameters to achieve higher group velocity, we expect it is possible to increase the propagation distance and thus the practical device size by an order of magnitude. There is also a limitation of the numerical aperture in the system, which is posed by the limited variability of the effective saturation magnetization (and therefore the refractive index) induced by irradiation. Although we demonstrated that a refractive index comparable to optics is achievable ($n\approx$~1.6), a higher change would improve the applicability and performance of magnonic devices. We see a potential improvement here, as we estimate that less than a third of the YIG film is affected by the FIB due to the shallow penetration depth. By using lighter ions, e.g. He$^+$, we expect that a stronger effect is achievable. Finally, a strong disturbance of the interference pattern is caused by the waves that are generated adjacent to the fabricated devices. This could be avoided by separating the device laterally from the neighboring structures, for which FIB irradiation could also be used, both for creating reflective and absorbing boundary conditions, exploiting the steeply increasing damping at moderately high doses.
\\
Magnonic systems themselves have limitations, perhaps the most important one is the nonzero damping that limits realizable device size and complexity. Thus, spin waves are not an ideal fit to replicate classical optical building blocks, they are more suitable for the approaches used in nanophotonics \cite{papp2021nanoscale, molesky2018inverse} -- we believe that our technique is also applicable to the realization of such structures. The automatic design of magnonic devices combined with a flexible and versatile fabrication method has the potential to raise the bar for magnonic device concepts and drive the field towards practical applications.
\\
In terms of applications, the 4f system (and similar constructions) carries the biggest potential. Based on the 4f system it is possible to realize a wide class of linear (Fourier domain) signal processing applications as it was well established in the field of optical computing \cite{goodman2005introduction}. Thus, successful implementation of 4f systems with spin waves may allow any linear signal processing task to be implemented in the magnonic domain. Such tasks are essential building blocks of neuromorphic computing pipelines and are in great demand for edge AI tasks.
\section*{Conclusion}
\label{conclusion}
Magnonics is often seen as an integration-friendly manifestation of photonics -- a technology that enables the chip-scale realization of wave-based, interference-based devices. The analogy between photonics and magnonics (as incomplete as it is) acts as a major driving force behind applications. Optical computing devices could directly be translated into the magnetic domain, enabling a number of computing and signal-processing applications.
\\
So far, however, the wave phenomena appearing in spin-wave optics remained a far cry from the complexity and sophistication of what is achievable in optics. To some extent, these challenges are technological: high-quality patterning of YIG is challenging. Moreover, patterning alone is insufficient to achieve an optics-like functionality: to steer waves, one needs to manipulate the index of refraction $n$, which requires the introduction of field or thickness gradients, beyond patterning.
\\
Our paper presents FIB irradiation of YIG as a straightforward technology to manipulate the index of refraction precisely and in a quasi-continuous way, enabling magnonic clones of optical components. In this work, we demonstrated various use cases where elements known from optics can be adapted in the spin-wave domain by using FIB. This direct-writing technology offers very flexible rapid prototyping, while it avoids many culprits of other patterning methods, e.g. resolution limitations, inability to produce gradients and defects on the patterned edges. We believe that the presented technology has a great potential to accelerate magnonic research due to its wide availability and easy adaptation. Although FIB itself is not applicable for mass production, the same devices could be straightforwardly mass-produced using the implanter technology omnipresent in industrial settings. Due to its flexibility, high resolution, wide availability and clear process-conversion for mass production, the presented technology has the potential to drive a 'magnonic revolution' and helps to bring spin-wave-based signal processors and computing accelerators to the market.
\bibliography{main}

\begin{thebibliography}{10}
\urlstyle{rm}
\expandafter\ifx\csname url\endcsname\relax
  \def\url#1{\texttt{#1}}\fi
\expandafter\ifx\csname urlprefix\endcsname\relax\def\urlprefix{URL }\fi
\expandafter\ifx\csname doiprefix\endcsname\relax\def\doiprefix{DOI: }\fi
\providecommand{\bibinfo}[2]{#2}
\providecommand{\eprint}[2][]{\url{#2}}

\bibitem{csaba2014spin}
\bibinfo{author}{Csaba, G.}, \bibinfo{author}{Papp, A.} \&
  \bibinfo{author}{Porod, W.}
\newblock \bibinfo{journal}{\bibinfo{title}{Spin-wave based realization of
  optical computing primitives}}.
\newblock {\emph{\JournalTitle{Journal of Applied Physics}}}
  \textbf{\bibinfo{volume}{115}}, \bibinfo{pages}{17C741}
  (\bibinfo{year}{2014}).

\bibitem{csaba2017perspectives}
\bibinfo{author}{Csaba, G.}, \bibinfo{author}{Papp, {\'A}.} \&
  \bibinfo{author}{Porod, W.}
\newblock \bibinfo{journal}{\bibinfo{title}{Perspectives of using spin waves
  for computing and signal processing}}.
\newblock {\emph{\JournalTitle{Physics Letters A}}}
  \textbf{\bibinfo{volume}{381}}, \bibinfo{pages}{1471--1476}
  (\bibinfo{year}{2017}).

\bibitem{wang2021inverse}
\bibinfo{author}{Wang, Q.}, \bibinfo{author}{Chumak, A.~V.} \&
  \bibinfo{author}{Pirro, P.}
\newblock \bibinfo{journal}{\bibinfo{title}{Inverse-design magnonic devices}}.
\newblock {\emph{\JournalTitle{Nature communications}}}
  \textbf{\bibinfo{volume}{12}}, \bibinfo{pages}{1--9} (\bibinfo{year}{2021}).

\bibitem{papp2021nanoscale}
\bibinfo{author}{Papp, {\'A}.}, \bibinfo{author}{Porod, W.} \&
  \bibinfo{author}{Csaba, G.}
\newblock \bibinfo{journal}{\bibinfo{title}{Nanoscale neural network using
  non-linear spin-wave interference}}.
\newblock {\emph{\JournalTitle{Nature communications}}}
  \textbf{\bibinfo{volume}{12}}, \bibinfo{pages}{1--8} (\bibinfo{year}{2021}).

\bibitem{papp2017nanoscale}
\bibinfo{author}{Papp, {\'A}.}, \bibinfo{author}{Porod, W.},
  \bibinfo{author}{Csurgay, {\'A}.~I.} \& \bibinfo{author}{Csaba, G.}
\newblock \bibinfo{journal}{\bibinfo{title}{Nanoscale spectrum analyzer based
  on spin-wave interference}}.
\newblock {\emph{\JournalTitle{Scientific reports}}}
  \textbf{\bibinfo{volume}{7}}, \bibinfo{pages}{1--9} (\bibinfo{year}{2017}).

\bibitem{zhu2017patterned}
\bibinfo{author}{Zhu, N.} \emph{et~al.}
\newblock \bibinfo{journal}{\bibinfo{title}{Patterned growth of crystalline
  y3fe5o12 nanostructures with engineered magnetic shape anisotropy}}.
\newblock {\emph{\JournalTitle{Applied Physics Letters}}}
  \textbf{\bibinfo{volume}{110}}, \bibinfo{pages}{252401}
  (\bibinfo{year}{2017}).

\bibitem{trempler2020integration}
\bibinfo{author}{Trempler, P.} \emph{et~al.}
\newblock \bibinfo{journal}{\bibinfo{title}{Integration and characterization of
  micron-sized yig structures with very low gilbert damping on arbitrary
  substrates}}.
\newblock {\emph{\JournalTitle{Applied Physics Letters}}}
  \textbf{\bibinfo{volume}{117}}, \bibinfo{pages}{232401}
  (\bibinfo{year}{2020}).

\bibitem{schlitz2019focused}
\bibinfo{author}{Schlitz, R.} \emph{et~al.}
\newblock \bibinfo{journal}{\bibinfo{title}{Focused ion beam modification of
  non-local magnon-based transport in yttrium iron garnet/platinum
  heterostructures}}.
\newblock {\emph{\JournalTitle{Applied Physics Letters}}}
  \textbf{\bibinfo{volume}{114}}, \bibinfo{pages}{252401}
  (\bibinfo{year}{2019}).

\bibitem{chumak2015magnon}
\bibinfo{author}{Chumak, A.~V.}, \bibinfo{author}{Vasyuchka, V.~I.},
  \bibinfo{author}{Serga, A.~A.} \& \bibinfo{author}{Hillebrands, B.}
\newblock \bibinfo{journal}{\bibinfo{title}{Magnon spintronics}}.
\newblock {\emph{\JournalTitle{Nature Physics}}} \textbf{\bibinfo{volume}{11}},
  \bibinfo{pages}{453--461} (\bibinfo{year}{2015}).

\bibitem{vogel2020optical}
\bibinfo{author}{Vogel, M.}, \bibinfo{author}{Pirro, P.},
  \bibinfo{author}{Hillebrands, B.} \& \bibinfo{author}{Von~Freymann, G.}
\newblock \bibinfo{journal}{\bibinfo{title}{Optical elements for anisotropic
  spin-wave propagation}}.
\newblock {\emph{\JournalTitle{Applied Physics Letters}}}
  \textbf{\bibinfo{volume}{116}}, \bibinfo{pages}{262404}
  (\bibinfo{year}{2020}).

\bibitem{albisetti2017nanopatterning}
\bibinfo{author}{Albisetti, E.} \emph{et~al.}
\newblock \bibinfo{journal}{\bibinfo{title}{Nanopatterning spin-textures: A
  route to reconfigurable magnonics}}.
\newblock {\emph{\JournalTitle{Aip Advances}}} \textbf{\bibinfo{volume}{7}},
  \bibinfo{pages}{055601} (\bibinfo{year}{2017}).

\bibitem{chappert1998planar}
\bibinfo{author}{Chappert, C.} \emph{et~al.}
\newblock \bibinfo{journal}{\bibinfo{title}{Planar patterned magnetic media
  obtained by ion irradiation}}.
\newblock {\emph{\JournalTitle{Science}}} \textbf{\bibinfo{volume}{280}},
  \bibinfo{pages}{1919--1922} (\bibinfo{year}{1998}).

\bibitem{mendsich2020perpendicular}
\bibinfo{author}{Mendsich, S.}, \bibinfo{author}{Ahrens, V.},
  \bibinfo{author}{Kiechle, M.}, \bibinfo{author}{Papp, A.} \&
  \bibinfo{author}{Becherer, M.}
\newblock \bibinfo{journal}{\bibinfo{title}{Perpendicular nanomagnetic logic
  based on low anisotropy co$\backslash$ni multilayer}}.
\newblock {\emph{\JournalTitle{Journal of Magnetism and Magnetic Materials}}}
  \textbf{\bibinfo{volume}{510}}, \bibinfo{pages}{166626}
  (\bibinfo{year}{2020}).

\bibitem{ruane2018controlling}
\bibinfo{author}{Ruane, W.} \emph{et~al.}
\newblock \bibinfo{journal}{\bibinfo{title}{Controlling and patterning the
  effective magnetization in y3fe5o12 thin films using ion irradiation}}.
\newblock {\emph{\JournalTitle{AIP Advances}}} \textbf{\bibinfo{volume}{8}},
  \bibinfo{pages}{056007} (\bibinfo{year}{2018}).

\bibitem{papp2016optically}
\bibinfo{author}{Papp, A.}, \bibinfo{author}{Csaba, G.} \&
  \bibinfo{author}{Porod, W.}
\newblock \bibinfo{title}{Optically-inspired computing based on spin waves}.
\newblock In \emph{\bibinfo{booktitle}{2016 IEEE International Conference on
  Rebooting Computing (ICRC)}}, \bibinfo{pages}{1--4}
  (\bibinfo{organization}{IEEE}, \bibinfo{year}{2016}).

\bibitem{davies2015towards}
\bibinfo{author}{Davies, C.~S.} \emph{et~al.}
\newblock \bibinfo{journal}{\bibinfo{title}{Towards graded-index magnonics:
  Steering spin waves in magnonic networks}}.
\newblock {\emph{\JournalTitle{Physical Review B}}}
  \textbf{\bibinfo{volume}{92}}, \bibinfo{pages}{020408}
  (\bibinfo{year}{2015}).

\bibitem{chang2018hybrid}
\bibinfo{author}{Chang, J.}, \bibinfo{author}{Sitzmann, V.},
  \bibinfo{author}{Dun, X.}, \bibinfo{author}{Heidrich, W.} \&
  \bibinfo{author}{Wetzstein, G.}
\newblock \bibinfo{journal}{\bibinfo{title}{Hybrid optical-electronic
  convolutional neural networks with optimized diffractive optics for image
  classification}}.
\newblock {\emph{\JournalTitle{Scientific Reports}}}
  \textbf{\bibinfo{volume}{8}}, \doiprefix\url{10.1038/s41598-018-30619-y}
  (\bibinfo{year}{2018}).

\bibitem{bauer2015nonlinear}
\bibinfo{author}{Bauer, H.~G.}, \bibinfo{author}{Majchrak, P.},
  \bibinfo{author}{Kachel, T.}, \bibinfo{author}{Back, C.~H.} \&
  \bibinfo{author}{Woltersdorf, G.}
\newblock \bibinfo{journal}{\bibinfo{title}{Nonlinear spin-wave excitations at
  low magnetic bias fields}}.
\newblock {\emph{\JournalTitle{Nature communications}}}
  \textbf{\bibinfo{volume}{6}}, \bibinfo{pages}{1--7} (\bibinfo{year}{2015}).

\bibitem{ziegler2013srim}
\bibinfo{author}{Ziegler, J.}
\newblock \bibinfo{journal}{\bibinfo{title}{Srim \& trim}}.
\newblock {\emph{\JournalTitle{http://www. srim. org/}}}
  (\bibinfo{year}{2013}).

\bibitem{papp2021experimental}
\bibinfo{author}{Papp, {\'A}.} \emph{et~al.}
\newblock \bibinfo{journal}{\bibinfo{title}{Experimental demonstration of a
  concave grating for spin waves in the rowland arrangement}}.
\newblock {\emph{\JournalTitle{Scientific Reports}}}
  \textbf{\bibinfo{volume}{11}}, \bibinfo{pages}{1--8} (\bibinfo{year}{2021}).

\bibitem{lenses2008}
\bibinfo{author}{Paschotta, R.}
\newblock \bibinfo{journal}{\bibinfo{title}{article on 'lenses' in the
  encyclopedia of laser physics and technology}}.
\newblock {\emph{\JournalTitle{1. edition, Wiley-VCH, ISBN 978-3-527-40828-3}}}
   (\bibinfo{year}{2008}).

\bibitem{ambs2010optical}
\bibinfo{author}{Ambs, P.}
\newblock \bibinfo{journal}{\bibinfo{title}{Optical computing: A 60-year
  adventure.}}
\newblock {\emph{\JournalTitle{Advances in Optical Technologies}}}
  (\bibinfo{year}{2010}).

\bibitem{goodman2005introduction}
\bibinfo{author}{Goodman, J.~W.}
\newblock \bibinfo{journal}{\bibinfo{title}{Introduction to fourier optics,
  roberts \& co}}.
\newblock {\emph{\JournalTitle{Publishers, Englewood, Colorado}}}
  (\bibinfo{year}{2005}).

\bibitem{davies2015graded}
\bibinfo{author}{Davies, C.~S.} \& \bibinfo{author}{Kruglyak, V.}
\newblock \bibinfo{journal}{\bibinfo{title}{Graded-index magnonics}}.
\newblock {\emph{\JournalTitle{Low Temperature Physics}}}
  \textbf{\bibinfo{volume}{41}}, \bibinfo{pages}{760--766}
  (\bibinfo{year}{2015}).

\bibitem{Smith2007modern}
\bibinfo{author}{Smith, W.~J.}
\newblock \emph{\bibinfo{title}{Modern Optical Engineering}}
  (\bibinfo{publisher}{McGraw- Hill Professional, 4th Edition},
  \bibinfo{year}{2007}).

\bibitem{molesky2018inverse}
\bibinfo{author}{Molesky, S.} \emph{et~al.}
\newblock \bibinfo{journal}{\bibinfo{title}{Inverse design in nanophotonics}}.
\newblock {\emph{\JournalTitle{Nature Photonics}}}
  \textbf{\bibinfo{volume}{12}}, \bibinfo{pages}{659--670}
  (\bibinfo{year}{2018}).

\bibitem{kiechle2019engineering}
\bibinfo{author}{Kiechle, M.} \& \bibinfo{author}{Mendisch, S.}
\newblock \bibinfo{title}{Engineering of sputter deposited yig-a comprehensive
  protocol for ultra-low damping magnetic thin films}.
\newblock In \emph{\bibinfo{booktitle}{Magnonics}} (\bibinfo{year}{2019}).

\bibitem{ding2020sputtering}
\bibinfo{author}{Ding, J.}, \bibinfo{author}{Liu, T.}, \bibinfo{author}{Chang,
  H.} \& \bibinfo{author}{Wu, M.}
\newblock \bibinfo{journal}{\bibinfo{title}{Sputtering growth of low-damping
  yttrium-iron-garnet thin films}}.
\newblock {\emph{\JournalTitle{IEEE Magnetics Letters}}}
  \textbf{\bibinfo{volume}{11}}, \bibinfo{pages}{1--6} (\bibinfo{year}{2020}).

\end{thebibliography}
\section*{Acknowledgements}
The authors want to thank all staff members and researchers working in the lab facilities of ZEITLab, 
and Tatyana Orlova and Maksym Zhukovski at the Imaging Facility at University of Notre Dame. 
Funding from the German Research Foundation (DFG), the German Academic Exchange Service (DAAD) and the Bavaria California Technology Center (BaCaTeC) is acknowledged. Adam Papp received funding from the PPD research program of the Hungarian Academy of Sciences.
\section*{Author contributions statement}
M. K.  A. P. and M. B. conceived the ideas and conducted the experiments.  A.P. designed and built the trMOKE microscope and advised in the measurements, 
M.K., A.P. Cs. Gy and M.B. wrote the manuscript. All authors analyzed the results and reviewed the manuscript.
\section*{Additional information}
\textbf{Competing interests:} The authors declare no competing interests. 
\section{Supplementary Material on Experimental Procedures}
\label{supplementary}
\subsection*{YIG film Fabrication and Characterization}
\label{sec:fabrication}
We use RF Magnetron sputter deposition to fabricate the renowned, damping-friendly configuration of Yttrium Iron Garnet ($Y_3I_5O_{12}$) thin films\cite{kiechle2019engineering} on Gadolinium Gallium Garnet (GGG) substrates from Saint-Gobain Crystals. For this, we had recourse to well-established procedures \cite{ding2020sputtering} and obtained best film qualities with a working pressure of 40 $\mu$Bar and an RF Power of 100 W, resulting in a deposition rate of 6.6 nm/min. Subsequently, recrystallization has been achieved with an annealing process in oxygen atmosphere at 700$^\circ$ C for 8 hours with a ramp up time of 10$^\circ$ C/h and a cool down of 1$^\circ$ C/h. The thickness $\mathrm{t_{YIG}}$ = 100 nm is chosen to maintain a stable and reproducible film quality across the experiments. The magnetic parameters of interest, effective magnetization and Gilbert damping, have been measured with broadband Ferromagnetic Resonance measurements. We obtain numerical values of $\mathrm{M_{eff}}$ between 110 and 120 kA/m and $\mathrm{\alpha_{YIG}}$ from 0.0005 to 0.0001 for the films used in our experiments. The physical film quality is certainly not as perfect as with LPE growth, especially at the interface to GGG (for details, we refer to Fig.~\ref{fig:TEM}. For the electrical excitation of spin waves (SWs), shorted co-planar microwave antennas with center conductor $\mathrm{s_{CPW}}$ and gap $\mathrm{g_{CPW}}$ widths ranging from 2 to 5 $\mu$m are fabricated on top of the YIG film with E-beam evaporated Aluminum (300-400 nm). 
\subsection*{Design and fabrication details of FIB structures}
\label{sec:design}
\begin{figure}[ht]
    \centering
    \vspace*{-0.2cm}
    \includegraphics[width=\textwidth]{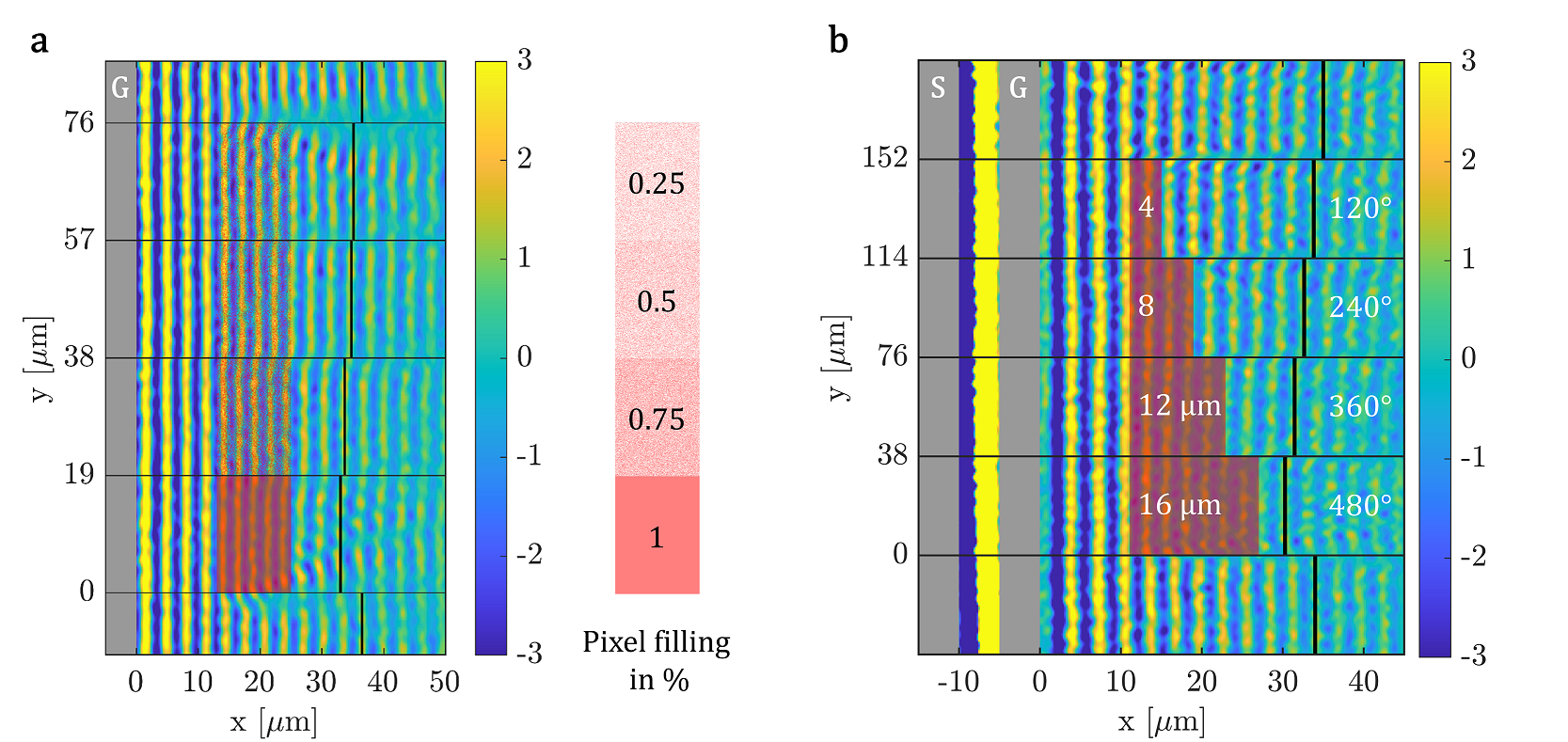}
    \caption{Dose and geometry driven $M_\mathrm{eff}$ patterning in YIG. (a) Demonstration of different filling factors and the resulting (small) wavelength change, highlighted by the indicated phase shift after the irradiation areas. (b) Phase change demonstration of spin waves propagation through irradiated areas of different lengths.}
    \label{fig:filling}
\end{figure}

Refractive index $n$ modifications can be achieved with low ion doses, and a convenient way to achieve smooth transitions of $n$ is changing the filling factor in pixel space of the FIB image. This way, only one global dose is applied, which means the current does not have to be changed pixel-wise. The effective ion dose is equivalent to the filling ratio presumend that the applied wavelength is much larger than the pixel size, e.g. a filling of 50$\%$ is half of the applied dose. Fig.~\ref{fig:filling}a shows an illustration of the effect. Generally, it is to mention that 50$\%$ filling does not mean 50$\%$ $M_\mathrm{eff}$ change (see Fig.~\ref{fig:dosemap}d) because of the non-linear relation. With this approach, we can realize magnetization landscapes with a resolution up to the minimum pixel size of the FIB image (10 nm). Furthermore, we can use the length of irradiated regions to create phase changes of desire, an illustration is shown in Fig.~\ref{fig:filling}b, and we used this example to choose the length for the phase plate in Fig.~\ref{fig:zoneplates}a (6 $\mu$m should be approximately 180
\subsection*{Ion Beam Irradiation Impact on Crystal Level}
\label{sec:TEM}
\begin{figure}[ht]
    \centering
    \includegraphics[width=\textwidth]{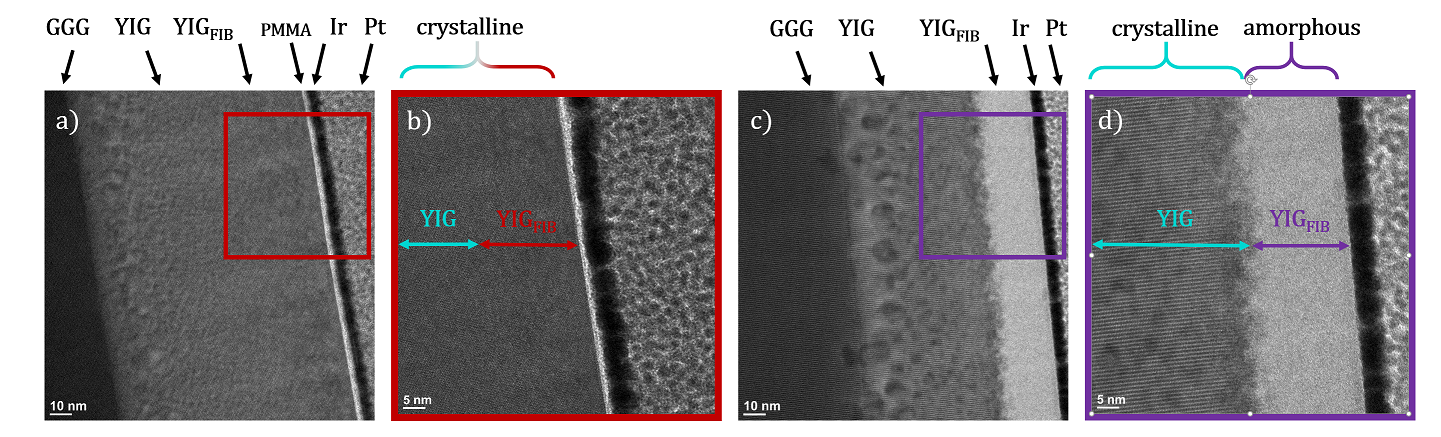}
    \caption{TEM images of FIB irradiated YIG thin films. (a) Cross-section of a YIG thin film irradiated at an ion dose of $1 \cdot 10^{13}$ ions/cm$^2$ and (b) Snippet of (a) showing the top part and the expected ion implantation depth indicated. The crystallinity of YIG is preserved. (c) Irradiation impact in YIG at a higher dose of $1.3 \cdot 10^{14}$ ions/cm$^2$ ions/cm2 and (d) Snippet of (c) revealing an amorphous top layer down to approximately 25\,nm.}
\label{fig:TEM}
\end{figure}

The average penetration depth of 50~keV accelerated Ga$^+$ ions in YIG is estimated to 24 nm according to TRIM simulations \cite{ziegler2013srim}, meaning only a part of the total film thickness (100\,nm) is affected by the ion irradiation. The observed magnetization change is a complex combination of multiple effects, mainly the anisotropy change due to dislocations in the YIG crystal structure leading to strain induced anisotropy, and the interaction with the interface to the underlying film part (and certainly with the underlying layer itself). 
To get insight on the structural properties of YIG, and more importantly on the physical influence of Ga$^+$ ion irradiation, we imaged ion irradiated thin films with transmission electron microscopy (TEM). For a dose regime of $10^{12}$ to the lower $10^{13}$ ions/cm$^2$, where $M_\mathrm{eff}$ is modified with only moderate increase in damping, there is no visible crystalline damage in YIG (Fig. \ref{fig:TEM}a,b). As for the interface to GGG, coarse structural bumps can be recognized, that could come from the high working pressure used during sputtering or an imperfect GGG interface, leading to a reorientation of growth direction. We neither think these are voids since the crystallinity inside of them is visible, nor that they have anything to do with the ion irradiation since it does not reach that far. We do notice that these defects appear even stronger in the sample shown in Fig. \ref{fig:TEM}c,d, which is most likely due to a skipped sputter cleaning step of the GGG. The ion dose is an order of magnitude higher ($10^{14}$ ions/cm$^2$), and spin waves are not detectable anymore with our measurement tools. The top layer of the 80\,nm thick YIG film has turned amorphous, whereby its thickness (25 to 28\,nm) is close to the expected ion implantation depth. In principle, spin waves should be able to travel underneath, but the uncontrolled interface of unknown width might just add to the degree of destruction.   
The demonstrated physical properties of the two ion dose regimes can be used for spin wave steering in different ways, i.e. the low dose for tuning the relative magnonic refractive index and the high dose to suppress spin wave propagation locally. Studying the underlying cause of the ion dose dependent change of $M_\mathrm{eff}$ deserves its own research and is beyond the scope of this work. After all, we find the effect FIB manipulation in YIG to be reversible by repeating a main fabrication step of YIG thin films: recrystallization via high temperature annealing (but impractical for device application).
\subsection*{Time-resolved 2D Optical Imaging of Spin Wave Patterns}
\label{sec:trMOKE}
To detect the dynamic in-plane magnetization components $\mathbf{m_x}$ and $\mathbf{m_y}$ in YIG, we use an in-house built longitudinal time-resolved magneto-optical Kerr effect microscope (trMOKE). The picosecond laser has a wavelength of 405 nm and pulses at 50 MHz and minimum step size of the scanning is 0.4 $\mu$m, limiting the resolution for spin waves to about 1 $\mu$m wavelength at about 10 GHz excitation frequency. With our samples we can detect spin waves amplitudes up to 180 $\mu$m away from the excitation antenna, the decay length is not only film/Gilbert damping-dependent but sensitive to the efficiency spectrum of the excitation antenna. The optical stage to guide the laser beam is led through a microscope with a low working distance 100x objective. For the bias field, one height-adjustable permanent magnet underneath the sample is used, as it provides a more stable and stronger field than a single-pole electromagnet, and the setup is optimized to image the magnetostatic forward volume configuration. Off-axis stray fields make the alignment normal to the sample cumbersome and can cause deviations in the expected focal distances and diffraction angles of the presented elements due to portions of anisotropic wave components. %
\end{document}